\documentclass[reprint, amsmath, amssymb, aps, prd]{revtex4-2}

\usepackage{graphicx}
\usepackage{dcolumn}
\usepackage{bm}
\usepackage{mathtools}
\usepackage[nice]{nicefrac}
\usepackage{graphicx}
\usepackage{caption}
\usepackage{xcolor}
\usepackage[bottom]{footmisc} 
\usepackage{subcaption}
\usepackage{ragged2e}
\usepackage{hyperref}

\newcommand{\capindent}{\justifying\leftskip=1em\rightskip=1em}

\newcommand{\dd}{\mathrm{d}}

\def\ps{\mathcal{P}_{_{\mathrm{S}}}}

\begin{document}

\title{Inverse-Scattering Reconstruction of Inflation from Scalar and Tensor Primordial Spectra}

\author{Jorge Mastache}%
\email{jh.mastache@mctp.mx}
\affiliation{Secretar\'ia de Ciencia, Humanidades, Tecnolog\'ia e Innovaci\'on, Av. Insurgentes Sur 1582, Colonia Cr\'edito
Constructor, Del. Benito Ju\'arez, 03940, Ciudad de M\'exico, M\'exico.}
\affiliation{Mesoamerican Centre for Theoretical Physics, Universidad Aut\'{o}noma de Chiapas,  Carretera Zapata Km. 4, Real del Bosque, 29040, Tuxtla Guti\'{e}rrez, Chiapas, M\'{e}xico.}

\author{Allan Hurtado}
\email{hur15011@uvg.edu.gt}
\affiliation{Instituto de Investigación en Ciencias Físicas y Matemáticas (IFIM), Escuela de Ciencias F\'isicas y Matem\'aticas, Universidad de San Carlos de Guatemala, Ciudad Universitaria Z.12, Guatemala}
\affiliation{Universidad del Valle de Guatemala, 18 Av. 11-95 Zona 15, Vista Hermosa III, Ciudad de Guatemala, Guatemala.}

\begin{abstract}
We develop an inverse-scattering framework to reconstruct the effective inflationary potentials governing scalar and tensor perturbations. By recasting the Mukhanov--Sasaki equation as a Schrödinger-like problem on the half-line, we identify the Bunch--Davies initial condition with the asymptotic Jost solution and show that the freeze-out amplitude of the growing mode is encoded in the corresponding Jost function. This allows the scalar and tensor primordial power spectra to be written in terms of $F^{(S)}_{\nu_s-1/2}(k)$ and $F^{(T)}_{\nu_t-1/2}(k)$, respectively, and leads to an inverse-scattering expression for the tensor-to-scalar ratio as a ratio of Jost amplitudes. We then test the reconstruction in the large-$k$ regime using the Born approximation, where the Marchenko equation becomes linear. As benchmarks, we consider a smooth quadratic potential and a step potential that transiently violates slow-roll and generates localized features in the primordial spectra. The reconstructed effective potentials reproduce the dominant behavior of both $z''/z$ and $a''/a$ for smooth slow-roll evolution, while localized discrepancies arise in the scalar sector when sharp features induce stronger scattering. Our results show that inverse scattering provides a physically transparent method for connecting features in the primordial spectra to the underlying inflationary dynamics, and that the Jost function acts as a sensitive diagnostic of departures from canonical slow-roll evolution.
\end{abstract}

\maketitle

\section{Introduction}\label{sec:introduction}
Inflation provides a compelling mechanism for generating the primordial scalar and tensor perturbations that seed the observed cosmic structure and leave imprints in the Cosmic Microwave Background (CMB) anisotropies \cite{Mukhanov:1985rz,Sasaki:1986hm,Bassett:2005xm,Martin:2013tda,Maldacena:2002vr}. In the standard canonical picture, the dynamics of these perturbations are governed by the Mukhanov--Sasaki equation, whose effective potentials are given by $z''/z$ for scalar perturbations and $a''/a$ for tensor perturbations. These quantities encode, respectively, the inflaton-background dynamics and the geometrical expansion history. Therefore, reconstructing them from the primordial power spectra provides a direct route to inferring the inflationary background.

Several approaches have been developed to reconstruct either the primordial power spectrum or the inflationary dynamics from cosmological data. These include penalized-likelihood and Bayesian methods, cubic-spline reconstructions, Richardson--Lucy deconvolution, generalized slow-roll approaches, Hamilton--Jacobi reconstruction, inflationary flow methods, and effective-field-theory-based reconstructions \cite{Planck:2018jri,Forconi:2021que,Wang:2024dtq,Liu:2024zql,Vazquez:2012ux,Lodha:2023jru,Kogo:2004vt,Hunt:2013bha,Hazra:2016fkm,Aich:2011qv,Guo:2011re,Bridle:2003sa,Gauthier:2012aq,Planck:2013jfk,Planck:2015sxf,Aslanyan:2014mqa,CORE:2016ymi,Shafieloo:2003gf,Hazra:2014jwa,Kadota:2005hv,Dvorkin:2009ne,Hu:2011vr,Leach:2005av,Dvorkin:2010dn,Dvorkin:2011ui,Cline:2006db,Bean:2008ga,Durakovic:2019kqq}. These methods are particularly relevant because present CMB anisotropy measurements probe the primordial spectrum over a limited range of scales, while future CMB spectral-distortion experiments may extend sensitivity to much smaller scales, potentially reaching $k\sim10^3$--$10^4\,{\rm Mpc}^{-1}$ \cite{Kogut:2011xw,PRISM:2013fvg,Chluba:2011hw,Chluba:2012gq,Khatri:2013dha,Abitbol:2017vwa,Chluba:2019nxa}. This extended window is especially important for testing inflationary scenarios that generate localized features or oscillatory structures in the primordial spectra.

Inverse scattering theory offers a complementary and non-parametric framework for this problem. Originally developed to reconstruct interaction potentials from scattering data \cite{newton2012inverse,chadan2012inverse}, it has found applications in several fields, including geophysics \cite{wang2016seismic} and cosmology \cite{Habib:2004hd,Mastache:2016ahe}. The key observation is that the Mukhanov--Sasaki equation can be recast as a Schrödinger-like equation on the half-line by identifying $r=-\tau$. In this mapping, the early-time Bunch--Davies condition corresponds to the large-$r$ plane-wave behavior of a Jost solution, while the late-time freeze-out amplitude is encoded in the small-$r$ behavior of the same solution.

The Jost function plays a central role in this construction. In a featureless slow-roll background, after the appropriate Hankel normalization, the Jost function carries only trivial scattering information. By contrast, transient violations of slow-roll, sharp features, or oscillatory modulations in the inflaton potential induce nontrivial $k$-dependence in the Jost function. Therefore, the Jost function provides a direct diagnostic of deviations from canonical slow-roll evolution. In the Marchenko approach, this information is used to construct the scattering kernel and reconstruct the effective potential. In the present cosmological setting, this means reconstructing $z''/z$ from the scalar primordial spectrum and $a''/a$ from the tensor spectrum.

Inflationary models with features provide a natural arena for testing this framework. Step-like potentials can generate transient departures from slow-roll and localized oscillatory features in the scalar primordial power spectrum \cite{Starobinsky:1992ts,Adams:2001vc,Palma:2017wxu}. Similar phenomenology can arise in models with periodic or drifting modulations, such as axion monodromy inflation, where oscillatory structures in the primordial spectrum are produced by modulations of the inflaton potential \cite{Silverstein:2008sg,McAllister:2008hb,Flauger:2009ab,Flauger:2014ana}. More general non-canonical scenarios, such as Dirac--Born--Infeld inflation, and models with nonlocal quantum corrections, also motivate reconstruction methods that are sensitive not only to the potential shape but also to departures from the simplest slow-roll dynamics \cite{Silverstein:2003hf,Woodard:2018gfj,Brooker:2017kjd,Brooker:2017kij}.

In this work, we develop an inverse-scattering formulation for both scalar and tensor inflationary perturbations. We express the primordial scalar and tensor spectra in terms of their corresponding Jost functions, $F^{(s)}_{\nu_s-1/2}(k)$ and $F^{(T)}_{\nu_t-1/2}(k)$, and derive the tensor-to-scalar ratio as a ratio of Jost amplitudes. We emphasize that the physically relevant cosmological mode is associated with the Jost solution, since it is fixed by the Bunch--Davies condition at early times and determines the growing freeze-out mode at late times. The regular solution, although standard in conventional inverse-scattering theory, corresponds to the decaying super-horizon branch when applied directly to the Mukhanov--Sasaki variable.

We then test the reconstruction in the large-$k$ regime using the Born approximation, which linearizes the Marchenko equation and provides a controlled weak-scattering limit. As benchmarks, we consider a smooth quadratic potential, which remains close to slow-roll, and a step potential, which transiently violates slow-roll and generates features in the primordial spectra. By comparing the reconstructed $z''/z$ and $a''/a$ with the simulated background quantities, we assess the accuracy and limitations of the Born-level inverse-scattering reconstruction.

The paper is organized as follows. In Section~\ref{sec:inflation_dynamic}, we review the background dynamics of canonical inflation and introduce scalar and tensor perturbations in a unified Mukhanov--Sasaki notation. In Section~\ref{sec:inverse_scattering}, we present the inverse-scattering framework, define the Jost function, and derive its connection with the scalar and tensor primordial spectra. We also discuss the tensor-to-scalar ratio in terms of Jost amplitudes and introduce a diagnostic variable for the relative evolution of scalar and tensor modes. In Section~\ref{sec:working_cases}, we apply the Born approximation to the quadratic and step potentials and compare the reconstructed effective potentials with the simulated background evolution. Finally, in Section~\ref{sec:conclusion}, we summarize our results and discuss future extensions beyond the Born approximation.

\section{Inflation and Perturbations} \label{sec:inflation_dynamic}
The inflationary phase of the Universe is driven by a scalar field, known as the inflaton, $\phi(t)$, evolving within a spatially flat, isotropic, and homogeneous Friedmann-Lemaître-Robertson-Walker (FLRW) background described by
\begin{equation}
\dd s^{2}= -\dd t^{2}+a(t)^{2}\delta_{ij}\dd x^{i}\dd x^{j},,
\label{eq:metric_frw}
\end{equation}
where $a(t)$ represents the scale factor. The dynamics of the inflaton and the cosmological expansion are described by the Friedmann and Klein-Gordon (KG) equations
\begin{eqnarray}
	&& H^{2} = \frac{\rho_{\phi}}{3M_{pl}^{2}} ,\label{eq:friedmann}\\
	&& \ddot{\phi}+3H\dot{\phi}+V_\phi'(\phi) = 0 , \label{eq:KG}
\end{eqnarray}
where dots denote derivatives with respect to cosmic time $t$, and primes denote derivatives with respect to the inflaton field, i.e., $V^\prime_{\phi} = dV(\phi)/d\phi$. Here, $H=\dot{a}/a$ is the Hubble parameter, and $M_{pl} = 1/\sqrt{8\pi G}$ is the reduced Planck mass. The energy density $\rho_\phi$ and pressure $P_\phi$ associated with the inflaton are defined through the energy-momentum tensor as 
\begin{equation}\label{4}
\rho_{\phi}=\frac{\dot{\phi}^{2}}{2}+V(\phi), \quad \quad P_\phi=\frac{\dot{\phi}^{2}}{2}-V(\phi).
\end{equation}
The number of e-folds, $N$, quantifies the amount of exponential expansion that occurs during inflation. It is defined as the logarithm of the ratio of the scale factor at the end of inflation ($a(t_e)$) to the scale factor at a time $t$, $a(t)$
\begin{equation}
    N \equiv \ln \frac{a(t_{e})}{a(t)} = \int_{t}^{t_{e}}H \dd t \; . 
    \label{eq:e_folds}
\end{equation}
Solving the cosmological horizon and flatness problems typically requires inflation to last for at least $N \approx 60$ e-folds. To facilitate numerical analyses and conceptual clarity, it is common to rewrite the inflationary equations using $N$ as the evolution variable, employing the relation $\dd N = H \dd t$ \cite{Liu:2010dh, Liu:2012iba}. In this notation, derivatives with respect to $N$ are indicated by the subscript $N$, such that for any function $f$ we have $f_N = df/dN$.

Inflation is sustained under slow-roll conditions, meaning that the kinetic energy of the inflaton is small compared to its potential energy. This condition is conveniently characterized by the first slow-roll parameter, $\epsilon$, defined as
\begin{equation}
    \epsilon = -\frac{\dot{H}}{H^2}=-\frac{H_N}{H}.
    \label{eq:epsilon}
\end{equation}
Inflation continues as long as $\epsilon < 1$. Another relevant slow-roll parameter, $\eta$, which measures the variation of $\epsilon$, is defined as
\begin{equation}\label{7}
\eta = \frac{\dot{\epsilon}}{H\epsilon} .
\end{equation}
Rewriting the KG equation, Eq.~\eqref{eq:KG}, in terms of $N$ yields:
\begin{equation}
    \phi_{N N}+ \left(\frac{H_N}{H}+3\right) \phi_N+\frac{1}{H^2} \frac{\partial V(\phi)}{\partial \phi}=0
    \label{eq:KG_f_N}
\end{equation}

Similarly, combining the Friedmann equation, Eq.\eqref{eq:friedmann}, with the continuity equation, $\dot{\rho}_\phi = -3H(\rho_\phi + P_\phi)$, we derive the following relation
\begin{equation}
	H_N = -\frac{H \phi_N^2}{2 M_{pl}^2}.
	\label{eq:friedmann_H_N}
\end{equation}
For a given potential $V(\phi)$, the coupled system formed by Eqs.~\eqref{eq:KG_f_N} and \eqref{eq:friedmann_H_N} can be solved numerically given suitable initial conditions. The inflationary period ends when the slow-roll condition $\epsilon =1$ is reached, thus constraining the initial conditions for $\phi$ and its derivatives. These initial conditions are typically chosen to ensure at least the minimal required number of e-folds, providing a complete description of the inflationary dynamics. 

\subsection{Scalar and Tensor Perturbations}

We analyze inflationary perturbations in terms of the curvature perturbation $\mathcal{R}$ and the tensor perturbation $\psi$. The corresponding scalar and tensor mode functions are expressed through the Mukhanov--Sasaki variables as
\begin{equation}\label{eq:mukhanov_variable}
	z \equiv \frac{a\dot{\phi}}{H}, 
	\qquad 
	u_k=-z\mathcal{R}_k,
	\qquad 
	v_k=a\psi_k \;,
\end{equation}
an approximation that will be used throughout this work is $z \simeq \sqrt{2\epsilon}\,a\,M_{\rm Pl}$, which follows from imposing the slow-roll approximation during inflation. The scalar and tensor perturbations satisfy the Mukhanov--Sasaki equation in Fourier space,
\begin{equation}\label{eq:mukhanov_sasaki}
    \mu_k''+\left(k^2-q(\tau)\right)\mu_k=0,
\end{equation}
where
\begin{equation}\label{eq:q_of_tau}
    \mu_k=
    \begin{cases}
        u_k & (\mathrm{scalar}),\\
        v_k & (\mathrm{tensor}),
    \end{cases}
    \qquad
    q(\tau)=
    \begin{cases} 
        z''/z & (\mathrm{scalar}),\\
        a''/a & (\mathrm{tensor}).
    \end{cases}
\end{equation}
Here, primes denote derivatives with respect to conformal time $\tau$, defined by $d\tau = dt/a$. 

If the effective potential satisfies    $q(\tau)=\frac{\nu^2-1/4}{\tau^2}$, the Mukhanov--Sasaki equation can be written as a Hankel equation of order $\nu$,
\begin{equation}\label{eq:mukhanov_sasaki_hankel}
    \mu_k''+ \left( k^2-\frac{\nu^2-1/4}{\tau^2} \right)\mu_k=0 \;,
\end{equation}
whose general solution is
\begin{equation}\label{hankel_solution}
    \mu_k= \frac{\sqrt{-\pi\tau}}{2}
    \left[ c_1(k)H_{\nu}^{(1)}(-k\tau) +c_2(k)H_{\nu}^{(2)}(-k\tau) \right].
\end{equation}
Here, $H_{\nu}^{(1)}$ and $H_{\nu}^{(2)}$ denote the Hankel functions of the first and second kind, respectively, of order $\nu$. This form is commonly obtained under the slow-roll approximation; for a generic inflationary background evolution, the mode equation does not admit an exact Hankel-function solution.

The perturbation dynamics change according to the wavelength relative to the comoving Hubble radius, $R_H = (aH)^{-1}$. For sub-Hubble scales ($k \gg aH$), the scalar Mukhanov-Sasaki equation simplifies to...
Initial conditions for these perturbations are imposed deep inside the Hubble radius (where $-k\tau \gg 1$, i.e, $k/aH \gg 1$ or similarly $-k\tau\rightarrow\infty$), under the assumption of the Bunch-Davies vacuum \cite{Bunch:1978yq}. In this regime, the sub-Hubble, the mode solutions. Therefore, $c_{1}(k) = 1$ and $c_{2}(k)=0$, this choice of constants (which represent amplitudes of the wave modes and normalize the solution) ensures that the solution behaves as a plane wave. The initial conditions at a given initial time $\tau_i = \tau(N_i)$ are
\begin{equation}\label{eq:initial_conditions}
	\mu_k(\tau_i) = \frac{1}{\sqrt{2k}} e^{-ik\tau_i}, \quad \mu_{k}(\tau_i) = -i \sqrt{\frac{k}{2}} \frac{1}{aH}e^{-ik\tau_i} \;.
\end{equation}

Conversely, for super-Hubble scales ($k \ll aH$, $k\tau \ll 1$), we use the small-argument expansion of the Hankel solution, Eq.\eqref{hankel_solution}, to the Mukhanov-Sasaki equation to get:
\begin{align}\label{eq:hankel_super_horizon}
    \mu_k(\tau) \simeq 
    &-\,i\,\frac{\Gamma(\nu)}{2\sqrt{\pi}} \left(\frac{k}{2}\right)^{-\nu} 
    (-\tau)^{\frac12-\nu} \nonumber\\
    &+ \frac{\sqrt{\pi}}{2\,\Gamma(\nu+1)} \left(\frac{k}{2}\right)^\nu 
    (-\tau)^{\frac12+\nu}.
\end{align}
This shows the two independent super-horizon modes, a growing dominant mode proportional to $(-\tau)^{1/2-\nu}$, and a decaying subdominant mode proportional to $(-\tau)^{1/2+\nu}$.

The primordial power spectra are computed directly from the mode functions evaluated at horizon crossing, corresponding to the time when the perturbations freeze out. Using the generalized notation introduced above, the primordial power spectrum can be written as
\begin{equation}\label{eq:PPS_generic}
    \mathcal{P}_{\alpha}(k,\tau)=
    \frac{k^3}{2\pi^2}
    \left|
    \frac{\mu_k(\tau)}{f(\tau)}
    \right|^2,
\end{equation}
where $\alpha=\{S,T\}$ denotes scalar and tensor perturbations, respectively, with $f(\tau)=z(\tau)$ for $\alpha=S$ and $f(\tau)=a(\tau)$ for $\alpha=T$. Thus, using the growing mode of Eq.~\eqref{eq:hankel_super_horizon} the primordial power spectrum for scalar pertubation will be 
\begin{align}\label{eq:power_spectrum_from_inflation}
    \ps(k) =\frac{2^{2 \nu-4} \Gamma^2(\nu)}{\pi^3} \frac{\left(1-\epsilon_1\right)^{2 \nu-1}}{\epsilon_1} \frac{H^2}{M_{p l}^2}\left(\frac{k}{a H}\right)^{3-2 \nu}
\end{align}
Note that for scalar perturbations, for $\nu=3/2$ one recovers the familiar nearly scale-invariant result.

To connect theoretical predictions with observations, the spectral indices associated with the generalized primordial power spectra, $\mathcal P_\alpha(k)$, are defined at a pivot scale ($k_* = 0.05\,\mathrm{Mpc}^{-1}$) as
\begin{equation}
	n_s -1 \equiv \frac{d\ln \mathcal P_S}{d\ln k}\bigg\vert_{k_*},
	\qquad
	n_T \equiv \frac{d\ln \mathcal P_T}{d\ln k}\bigg\vert_{k_*}.
\end{equation}

\begin{figure*}[t]
    \centering
    \begin{subfigure}[t]{0.48\textwidth}
        \centering
        \includegraphics[width=\linewidth]{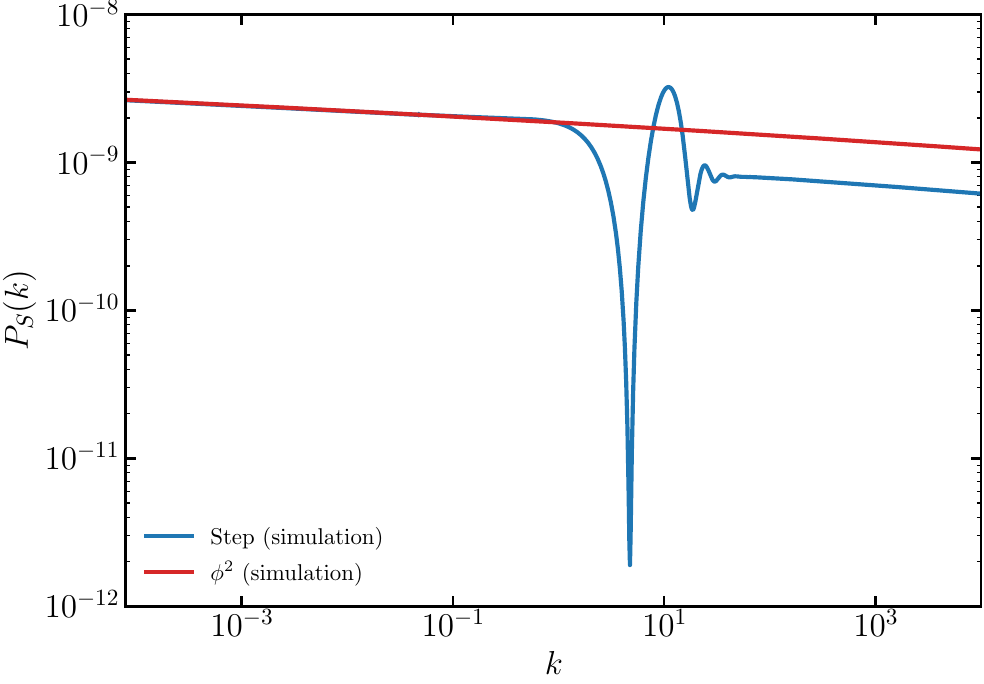}
        \caption{Scalar primordial power spectrum $P_S(k)$.}
    \end{subfigure}
    \hfill
    \begin{subfigure}[t]{0.48\textwidth}
        \centering
        \includegraphics[width=\linewidth]{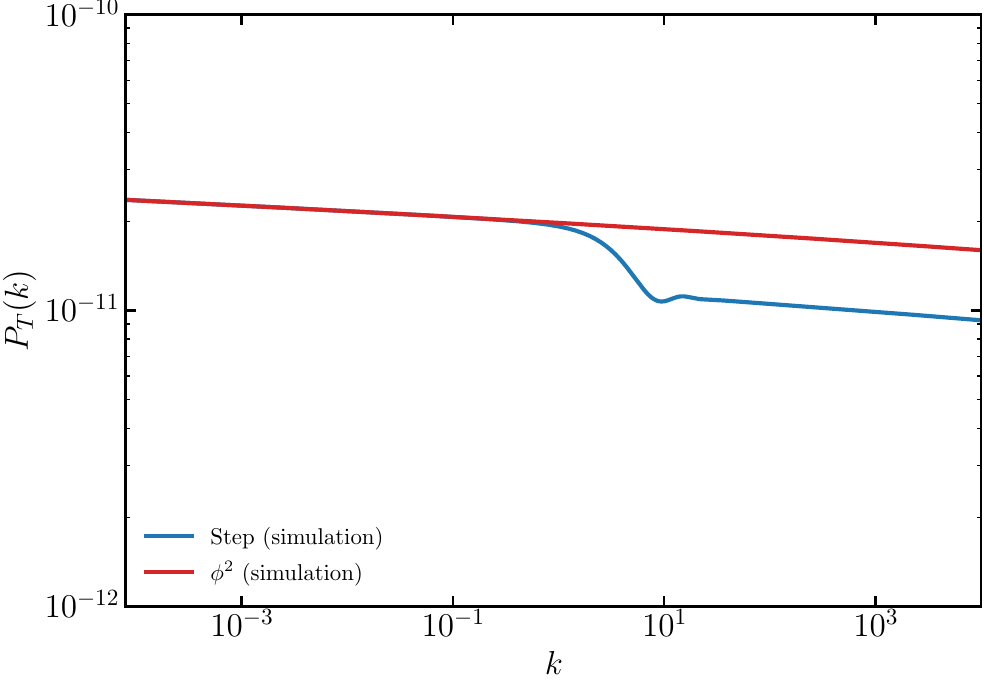}
        \caption{Tensor primordial power spectrum $P_T(k)$.}
    \end{subfigure}

    \caption{
    \capindent
    Scalar (left) and tensor (right) primordial power spectra obtained from numerical solutions of the Mukhanov--Sasaki equation for a quadratic potential ($\phi^2$) and a step potential. 
    The $\phi^2$ model yields an approximately scale-invariant spectrum, while the step potential introduces localized oscillatory features associated with transient deviations from slow-roll evolution.
    }
    \label{fig:pps_scalar_tensor}
\end{figure*}

\section{Inverse-Scattering and the Jost function} \label{sec:inverse_scattering} 
\subsection{Mathematical Framework}\label{subsec:mathematical_framework}

\noindent Inverse scattering theory originates from quantum mechanics, specifically scattering problems. While direct scattering aims to determine the scattered wavefunction given a potential, inverse scattering seeks to reconstruct the potential from observed scattering data.

\noindent The starting point is the reduced radial Schrödinger equation 

\begin{equation} \label{eq:schrodinger_radial}
\frac{d^{2}}{dr^{2}}\psi_{\ell}(k,r)+\left(k^{2}-\frac{\ell(\ell+1)}{r^{2}}\right)\psi_{\ell}(k,r)=V_{s}(r)\psi_{\ell}(k,r) \, ,
\end{equation} 

\noindent where $V_s(r)$ is the potential, assumed real and locally integrable (in the Lebesgue sense \cite{chadan2012inverse}), satisfying the following conditions near the origin and infinity 

\begin{align}
&\int_{0}^{a}r|V_{s}(r)|dr<\infty, \quad a<\infty \, , \label{eq:potential_origin} \\
&\int_{b}^{\infty}|V_{s}(r)|rdr<\infty, \quad b\geq 0 \, . \label{eq:potential_infty}
\end{align}

Condition \eqref{eq:potential_infty} implies that the potential is regular, decreasing more rapidly than $r^{-2}$ at infinity, having a finite number of bound states, and ensuring a well-defined inverse scattering transform. Singular potentials violate this condition.

In general, Schrödinger’s equation is singular at $r=0$. For the free potential case ($V_s=0$), the regular solution ($\varphi_\ell(k,r)$) at the origin is given by 
\begin{equation}\label{eq:free_solution}
	\varphi_{\ell}(kr)=\left(\frac{\pi kr}{2}\right)^{1/2}J_{\ell+1/2}(kr) \,,
\end{equation}
where $J_{\ell}$ is a Bessel function. Its asymptotic behaviors are
\begin{align}
\lim_{r\to 0} \varphi_{\ell}(kr) &\simeq \frac{(kr)^{\ell+1}}{(2\ell+1)!!} \,, \label{eq:free_solution_origin}
\end{align}
where the double factorial $(2\ell+1)!!$ is defined conventionally. It will be useful to keep in mind the following identity $(2\ell-1)!!=2^{\ell}\Gamma(\frac{1}{2}+\ell)/\Gamma(\frac{1}{2})$.
In the presence of interactions ($V_s\neq 0$), we assume that the regular solution satisfying the same boundary condition of the last equation, 
$\varphi_{\ell}(k,r)\propto r^{\ell+1}$.

Physically, the wavefunction $\psi_{\ell}$ must vanish at the origin; thus we introduce the free outgoing wave solution
\begin{equation}\label{eq:free_outgoing}
	w_{\ell}(kr)=i e^{i\pi \ell}\left(\frac{\pi}{2}kr\right)^{1/2}H_{\ell+1/2}^{(1)}(kr) \,,
\end{equation}
with $H_{\ell}^{(1)}$ is the Hankel function of the first kind of order $\ell$. Its asymptotic behaviors are
\begin{align}
	\lim_{r\to 0} w_{\ell}(kr)&\simeq (-1)^{\ell}(2\ell-1)!!(kr)^{-\ell} \,, \label{eq:free_outgoing_origin} \\
	\lim_{r\to\infty} w_{\ell}(kr)&\simeq e^{i\pi \ell/2}e^{ikr} \,, \label{eq:free_outgoing_infty}
\end{align}
this is, at infinity, it behaves as a plane wave, and at the origin is singular since $w_\ell \propto r^{-\ell}$. This behaves like a plane wave at large $r$ and corresponds to the subhorizon Bunch–Davies-like oscillatory regime, as we previously saw.

We now define the Jost solution ($f_{\ell}(k,r)$) to the Schrodinger equation that satisfy the asymptotic conditions 
\begin{eqnarray}\label{eq:jost_solution}
	\lim_{r\to\infty} f_{\ell}(k,r) &\propto& e^{ikr},. \\
    \lim_{r\to 0} f_\ell(k,r) &\sim& A_\ell(k)\, r^{-\ell} + B_\ell(k)\, r^{\ell+1}\,,
\end{eqnarray}
where the $r^{-\ell}$ corresponds to the singular solution Eq.\eqref{eq:free_outgoing_origin}, while the $r^{\ell+1}$ term is the regular contribution, Eq.\eqref{eq:free_solution_origin}. The ingoing and outgoing solutions $f_{\ell}(\pm k,r)$ are linearly independent with Wronskian. 
\begin{equation}\label{eq:jost_wronskian}
	W\{f_{\ell}(k,r),f_{\ell}(-k,r)\}=(-1)^{\ell+1}2ik,.
\end{equation}

The regular solution can be expressed as the following linear combination
\begin{equation}\label{eq:regular_solution}
	\varphi_{\ell}=\frac{i}{2k^{\ell+1}}\left[F_{\ell}(k)f_{\ell}(-k,r)+(-1)^{\ell}F_{\ell}(-k)f_{\ell}(k,r)\right],
\end{equation}
where $F_{\ell}(k)$ is the Jost function, defined through the Wronskian 
\begin{equation}\label{eq:jost_function}
	F_{\ell}(k)=(-k)^{\ell}W\{ f_{\ell}(k,r),\varphi_{\ell}(k,r)\} \, .
\end{equation}

The Jost function is normalized taken as the Bessel solution for the free potential, Eq.\eqref{eq:free_solution_origin}
\begin{equation}\label{eq:jost_normalized}
	F_{\ell}(k) \propto \lim_{r\to 0}(2\ell+1)!!r^{\ell}f_{\ell}(k,r) \, ,
\end{equation}
this guarantees that in the free case, $F_\ell(k) = 1$. To complete the definition, one rescales $f_\ell(k,r)$ by removing its singular behavior and normalize it to match the regular solution, therefore
\begin{equation} \label{eq:jost_function_origin}
    F_{\ell}(k)=\lim_{r\to 0}\left[\frac{e^{-i\pi \ell}\Gamma\left(\frac{1}{2}\right)}{\Gamma\left(\ell+\frac{1}{2}\right)}\left(\frac{kr}{2}\right)^{\ell}f_{\ell}(k,r)\right]\,.
\end{equation}
This combination assure a constant behavior when $r \to 0$, since
$f_\ell(k,r) \sim r^{-\ell}$ and extract the coefficient that encodes the full scattering information. In the cosmological context, it encodes how features of the effective potential (hence, inflationary dynamics) map into observables such as the primordial power spectrum.

Once $F_\ell(k)$ is known, one can construct the scattering matrix
\begin{equation}
    S_\ell(k)=\frac{F_\ell(-k)}{F_\ell(k)} =\frac{F_\ell^*(k)}{F_\ell(k)} =e^{2i\delta_\ell(k)},
\end{equation}
so the Jost function determines the full scattering data on the continuous spectrum. If bound states exist, their discrete data must also be included, but in most inflationary applications, one usually works with the continuum only.

If $F_\ell(k)$ is analytic in the upper half-plane, has no zeros there, and satisfies the usual asymptotic normalization $ F_\ell(k)\to 1$ as $k\to\infty$, then its phase is determined by its modulus through a dispersion relation. A convenient form is
\begin{equation}    
    \delta_\ell(k) = \frac{2k}{\pi}\,\mathcal P\!\!\int_0^\infty \frac{\ln |F_\ell(k')|}{k'^2-k^2}\,dk' \;.
\end{equation}
where $\mathcal P$ denotes the Cauchy principal value of the integral.

The key point is that the Marchenko kernel is built from the scattering data. Since $S_\ell(k)$ is determined by $F_\ell(k)$, this means that the Jost function fixes the Marchenko input kernel. In partial-wave form, one defines an input kernel $\mathcal F_\ell(r,t)$ through an integral of the schematic form
\begin{equation}\label{eq:input_kernel}
    \mathcal F_\ell(r,t) = \frac{1}{2\pi}\int_{-\infty}^{\infty} \left[1-S_\ell(k)\right] h_\ell^{(+)}(kr)\,h_\ell^{(+)}(kt)\,dk
\end{equation}
where $h_\ell^{(+)}$ denotes the outgoing Riccati–Hankel function. Then one solves the Marchenko integral equation
\begin{equation}
    A_\ell(r,t)+\mathcal F_\ell(r,t)
+\int_r^\infty A_\ell(r,s)\,\mathcal F_\ell(s,t)\,ds=0,
\qquad t\ge r.    
\end{equation}
Solving the Marchenko integral equation yields the transformation kernel $A_\ell(r,t)$, from which the reconstructed potential is obtained through 
\begin{equation}
    V_s(r)=-2\frac{d}{dr}A_\ell(r,r)
\end{equation}
Thus, the Jost function provides the bridge between the observable spectrum and the reconstructed effective potential $V_s(r)$, and hence to the inflationary background encoded in $z''/z$ or $a''/a$.

\subsection{Jost Function Analysis}   \label{subsec: Jost_function_analysis} 
Connecting the inverse-scattering formalism to cosmological perturbations we introduces a radial-like coordinate $r$, with minus conformal time, $r=-\eta$.  When $\tau\to -\infty$ marks the beginning of inflation (large $r$), while $\tau\to 0^-$ corresponds to the end of inflation (small $r$). The observable primordial spectrum is determined not by this early-time oscillatory limit, but by the late-time behavior of the modes, namely the $r\to 0$ limit. For this reason, one must connect the Jost solution (Bunch-Davis condition) at infinity to its asymptotic form near the origin.

First, we identify that the effective potential is related to
\begin{equation}\label{22}
    V_{s}(r)=\frac{z''}{z}-\frac{\ell(\ell+1)}{r^2} \,,
\end{equation}
here, the term $\ell(\ell+1)/r^2$ captures the leading near-origin behavior, $r\to 0$, corresponding to late times, that is, to the superhorizon regime. This implies the relation between the order $\ell$ in quantum mechanics and the parameter $\nu$ in inverse scattering, namely, $\ell=\nu-\frac12$.

The Bunch-Davis sub-horizon condition for the solution $\mu_{k}$ is characterized by an oscillatory behavior, which is equivalent to the Jost solution, Eq.\eqref{eq:jost_solution}, at $r\rightarrow \infty$, defined by its plane-wave behavior at large r; both approaches share the same initial condition, therefore one identifies $\mu_k(r)=f_\ell(k,r)/\sqrt{2k}$ for the positive-frequency mode.

The inverse-scattering formulation developed in this work applies analogously to both scalar and tensor perturbations. Keeping use of the unified notation with $\alpha=\{s,t\}$ labels the scalar and tensor sectors, respectively. Similarly, the Hankel index is denoted by $\nu_\alpha$, with
\begin{equation}\label{eq:scalar_tensor_index}
    \nu_s=\frac32+\epsilon+\frac{\eta}{2}, \qquad \nu_t=\frac32+\epsilon.
\end{equation}
the small-$r$ asymptotic behavior of the Jost solution implies that the freeze-out amplitude of the Mukhanov--Sasaki variable is proportional to the corresponding Jost function,
\begin{equation}\label{eq:generic_jost_growing_mode}
    \mu_{\alpha,k}(r) \underset{r\rightarrow0}{\longrightarrow} e^{i\pi(\nu_\alpha-\frac12)} \frac{\Gamma(\nu_\alpha)}{\sqrt{\pi}} 2^{\nu_\alpha-1}k^{-\nu_\alpha}(-\tau)^{\frac12-\nu_\alpha}F^{(\alpha)}_{\nu_\alpha-\frac12}(k).
\end{equation}
Equation~\eqref{eq:generic_jost_growing_mode} corresponds to the Jost-function representation of the freeze-out solution obtained from the Hankel expansion of the Mukhanov--Sasaki equation. In the exact Hankel background, the Jost function reduces to a pure phase. More generally, all departures from exact slow-roll evolution are encoded in the structure of the Jost function $F^{(\alpha)}_{\nu_\alpha-\frac12}(k)$.

The reconstruction can also be formulated in terms of the regular solution. Since the regular and Jost solutions are related through Eq.~\eqref{eq:regular_solution}, the Jost function acts as the coefficient connecting the solution normalized at the origin with the plane-wave solutions normalized at infinity. Using the asymptotic behavior $f_\ell(\pm k,r)\sim e^{\mp ikr}$, the regular solution near the origin behaves as
\begin{equation}\label{eq:generic_regular_solution}
    \mu_{\alpha}(k,r) \underset{r\rightarrow0}{\longrightarrow}\frac{\sqrt{2}\,k^{\nu_\alpha}i}{F^{(\alpha)}_{\nu_\alpha-\frac12}(-k)}\frac{1}{(2\ell_\alpha+1)!!} (-\tau)^{\nu_\alpha+\frac12}.
\end{equation}
This asymptotic behavior is proportional to the decaying mode of the Hankel solution, $\mu_{\alpha,k}(\tau)\propto(-\tau)^{\nu_\alpha+\frac12}$, while the physical freeze-out mode corresponds to the growing solution $\mu_{\alpha,k}(\tau)\propto(-\tau)^{\frac12-\nu_\alpha}$.

Consequently, the physical cosmological perturbation is associated with the Jost solution rather than the regular solution. In this sense, the small-$r$ behavior of the Jost function determines the observable freeze-out amplitude and therefore the primordial spectrum.

The primordial power spectrum can then be written generically as
\begin{equation}\label{eq:generic_jost_ps}
    \mathcal P_\alpha(k) \propto k^{3-2\nu_\alpha} \left| F^{(\alpha)}_{\nu_\alpha-\frac12}(k) \right|^2,
\end{equation}
up to normalization factors that depend on the scalar or tensor sector.

\subsubsection{Scalar modes} \label{subsubsec:scalar_modes}
We now specialize in scalar perturbations. During slow-roll inflation, and neglecting higher-order contributions $\mathcal O(\epsilon^2,\eta^2)$, using the scalar Mukhanov variable $z=\frac{a\dot\phi}{H}$ and using the relations $\epsilon'=\eta\mathcal H\epsilon$ and $\mathcal H'=\mathcal H^2(1-\epsilon)$ the effective scalar potential becomes
\begin{equation}\label{eq:scalar_effective_potential}
    \frac{z''}{z} = \frac{1}{\tau^2} \left( 2+3\epsilon+\frac32\eta \right)
\end{equation}
with $\nu_s$, in Eq.~\eqref{eq:scalar_tensor_index}, is approximately constant during slow-roll inflation.

The super-Hubble growing mode therefore behaves as $z(\tau)\propto(-\tau)^{\frac12-\nu_S}$,
and the scalar spectral index satisfies
\begin{equation}
    n_s - 1 = 3-2\nu_s = -2\epsilon-\eta.
\end{equation}

Substituting the scalar Jost asymptotic solution, Eq.~\eqref{eq:generic_jost_growing_mode}, into the definition of the scalar primordial spectrum, Eq.~\eqref{eq:PPS_generic}, yields
\begin{equation}\label{eq:scalar_jost_ps}
    \mathcal P_s(k) = \frac{2^{2\nu_s-4}\Gamma^2(\nu_s)} {\pi^3} \frac{(1-\epsilon_1)^{2\nu_s-1}} {\epsilon_1} \frac{H^2}{M_{\rm Pl}^2} \left( \frac{k}{aH} \right)^{3-2\nu_s} \left| F^{(S)}_{\nu_s-\frac12}(k) \right|^2.
\end{equation}
Comparing Eq.~\eqref{eq:power_spectrum_from_inflation} with Eq.~\eqref{eq:scalar_jost_ps} tell us that all departures from exact slow-roll evolution are encoded in the scalar Jost function. Unlike the tensor sector, scalar perturbations probe the combined inflaton-background dynamics through the effective potential $z''/z$. We write the power spectrum coming from the regular solution, Eq.~\eqref{eq:generic_regular_solution} in Appendix~\ref{appendix:regular_solution} keeping in mind that it correspond to the decaying super-horizon mode.

\subsubsection{Tensorial modes}\label{subsubsec: tensorial_modes}   
We now specialize to tensor perturbations. In this sector, the Mukhanov--Sasaki equation is governed by the effective potential $a''/a$, rather than by the scalar combination $z''/z$. Therefore, tensor modes probe directly the background expansion history. During slow-roll inflation, and neglecting higher-order contributions $\mathcal O(\epsilon^2,\eta^2)$, taking two conformal-time derivatives $a'' = a\left(\mathcal H^2+\mathcal H'\right)$, using $\mathcal H'=\mathcal H^2(1-\epsilon)$ we obtain to first order
\begin{equation}\label{eq:tensor_effective_potential}
    \frac{a''}{a} = \frac{1}{\tau^2} \left(2+3\epsilon\right),
\end{equation}
where $\nu_t$, defined in Eq.~\eqref{eq:scalar_tensor_index}, is approximately constant during slow-roll inflation. The tensor spectral index is then
\begin{equation}
    n_t = 3-2\nu_t = -2\epsilon.
\end{equation}
Using Eq.~\eqref{eq:generic_jost_growing_mode}, the tensor Jost function determines the freeze-out amplitude of the gravitational-wave perturbation. Therefore, the tensor primordial spectrum, Eq.~\eqref{eq:PPS_generic}, yields
\begin{equation}\label{eq:tensor_jost_ps_H}
    \mathcal P_t(k) = \frac{2^{2\nu_t-2}\Gamma^2(\nu_t) (1-\epsilon_1)^{2\nu_t-1}} {\pi^3} \frac{H^2} {M_{\rm Pl}^2} \left( \frac{k}{aH} \right)^{3-2\nu_t} \left| F^{(T)}_{\nu_t-\frac12}(k) \right|^2.
\end{equation}

Equation~\eqref{eq:tensor_jost_ps_H} is the tensor analogue of Eq.~\eqref{eq:scalar_jost_ps}. The main structural difference is the absence of the factor $1/\epsilon_1$, which appears only in the scalar spectrum through $z\simeq a\sqrt{2\epsilon_1}\,M_{\rm Pl}$. Thus, while scalar inverse scattering reconstructs the effective potential $z''/z$, tensor inverse scattering reconstructs the geometric potential $a''/a$ directly. As in the scalar case, departures from an exact Hankel background are encoded in the tensor Jost function $F^{(T)}_{\nu_t-\frac12}(k)$.

\subsubsection{Tensor-to-scalar ratio} \label{subsubsec:tensor_to_scalar_ratio}
The tensor-to-scalar ratio is defined as the ratio between the tensor and scalar primordial power spectra,
\begin{equation}\label{eq:r_definition_jost}
    \mathbf r(k) \equiv \frac{\mathcal P_t(k)} {\mathcal P_s(k)}.
\end{equation}
This quantity provides a direct measure of the relative amplitude of primordial gravitational waves with respect to scalar curvature perturbations and constitutes one of the primary observational probes of inflationary dynamics.

Using the inverse-scattering expressions for the scalar and tensor power spectra, Eqs.~\eqref{eq:scalar_jost_ps} and \eqref{eq:tensor_jost_ps_H}, the tensor-to-scalar ratio becomes
\begin{align}\label{eq:r_jost_general}
    \mathbf r(k) = 4\frac{\Gamma^2(\nu_t)} {\Gamma^2(\nu_s)} \frac{\epsilon_1}{\left(1-\epsilon_1\right)^{2(\nu_s-\nu_t)}}  \left(\frac{k}{aH}\right)^{2(\nu_s-\nu_t)} \frac{ \left|F^{(T)}_{\nu_t-\frac12}(k)\right|^2 }{ \left|F^{(S)}_{\nu_s-\frac12}(k)\right|^2 }.
\end{align}
Equation~\eqref{eq:r_jost_general} shows that the tensor-to-scalar ratio can be interpreted as the relative scattering amplitude between tensor and scalar perturbations. In particular, deviations from canonical slow-roll evolution are encoded in the ratio of tensor and scalar Jost functions,
\[
    \frac{ F^{(T)}_{\nu_t-\frac12}(k) }{ F^{(S)}_{\nu_s-\frac12}(k) }.
\]
For an exact Hankel background, both Jost functions reduce to pure phases, and the standard slow-roll result is recovered.

To leading order in slow-roll, $\epsilon \ll1$, $\eta\ll1$, so that $\Gamma(\nu_s)\simeq\Gamma(\nu_t)\simeq\Gamma\left(\frac32\right)$, and evaluating at horizon crossing, $k=aH$, and neglecting higher-order corrections, Eq.~\eqref{eq:r_jost_general} reduces to
\begin{equation}\label{eq:r_slow_roll}
    \mathbf r \simeq 16\epsilon.
\end{equation}
Using the consistency relation for single-field slow-roll inflation, $n_t=-2\epsilon$,
one finally obtains $\mathbf r=-8n_t$. For completeness, Appendix~\ref{appendix:additional_relations} shows how the leading-order slow-roll relation can be formally written in terms of the effective potential $z''/z$ evaluated at horizon crossing.

The inverse-scattering formalism also motivates the introduction of a diagnostic variable defined as the ratio between tensor and scalar Mukhanov--Sasaki modes,
\begin{equation}\label{eq:Xk_definition}
    X_k(\tau) \equiv \frac{v_k(\tau)} {u_k(\tau)}.
\end{equation}
The quantity $X_k$ can be interpreted as the amplitude-level analogue of the tensor-to-scalar ratio since $\mathbf r(k,\tau) = 2\epsilon(\tau)M_{\rm Pl}^2 |X_k(\tau)|^2$. Substituting Eq.~\eqref{eq:Xk_definition} into the scalar and tensor Mukhanov--Sasaki equations yields
\begin{equation}\label{eq:Xk_equation}
    X_k''+2\frac{u_k'}{u_k}X_k'+\left(\frac{z''}{z}-\frac{a''}{a} \right)X_k = 0.
\end{equation}
This equation governs the relative evolution between tensor and scalar perturbations and depends explicitly on the difference between the scalar and tensor effective potentials. In contrast to the observable tensor-to-scalar ratio, which is defined only after freeze-out, the variable $X_k$ retains the full dynamical information of the perturbation evolution. Consequently, Eq.~\eqref{eq:Xk_equation} provides an alternative perspective on the tensor-to-scalar ratio within the inverse-scattering framework, relating the evolution of the perturbation ratio directly to the reconstructed effective potentials. A logarithmic formulation of Eq.~\eqref{eq:Xk_equation} in terms of $Y_k \equiv \ln X_k$, which leads to a Ricatti-type equation and may be useful for future analyses, is provided in Appendix~\ref{appendix:additional_relations}.

The equation also admits a natural Schrödinger/WKB interpretation. The coefficient $2 u_k'/u_k$ acts as an effective friction term determined by the scalar background solution, while the quantity $(\frac{z''}{z}-\frac{a''}{a})$ plays the role of an effective interaction potential controlling the departure between scalar and tensor evolution. This structure allows analytical control of Eq.~\eqref{eq:Xk_equation} in both the sub-horizon and super-horizon regimes.

Deep inside the horizon, where $k^2\gg z''/z,a''/a$, the scalar mode behaves as $u_k(\tau) \propto e^{-ik\tau}$, implying $\frac{u_k'}{u_k}\simeq -ik$. Equation~\eqref{eq:Xk_equation} then reduces to
\begin{equation}\label{eq:Xk_subhorizon}
    X_k'' - 2ikX_k' 0,
\end{equation}
whose solution is $ X_k(\tau) = C_1 + C_2 e^{2ik\tau}$. Therefore, at early times, the tensor-to-scalar evolution variable rapidly oscillates, reflecting the fact that scalar and tensor perturbations share the same Bunch--Davies vacuum structure in the sub-horizon regime.

Conversely, in the super-horizon regime, where $-k\tau\ll1$, the growing scalar mode satisfies $u_k(\tau)\propto (-\tau)^{\frac12-\nu_s}$, such that
\begin{equation}
    \frac{u_k'}{u_k} \simeq - \frac{1}{\tau}\left( \frac12-\nu_s\right) = -\frac{1}{\tau}\left(1+\epsilon+\frac{\eta}{2}\right).
\end{equation}
Equation~\eqref{eq:Xk_equation} becomes
\begin{equation}\label{eq:Xk_superhorizon}
    X_k'' - \frac{2+2\epsilon+\eta}{\tau}X_k' + \left( \frac{z''}{z} - \frac{a''}{a} \right)X_k = 0.
\end{equation}
The super-horizon evolution is therefore sourced entirely by the difference between the scalar and tensor effective potentials, which vanishes in exact de Sitter spacetime.

Indeed, in the exact de Sitter limit, $\epsilon=\eta=0$, implies $\frac{z''}{z} = \frac{a''}{a} = \frac{2}{\tau^2}$, and Eq.~\eqref{eq:Xk_superhorizon} simplifies to
\begin{equation}\label{eq:Xk_desitter}
    X_k'' - \frac{2}{\tau}X_k' = 0,
\end{equation}
whose exact solution is $X_k(\tau) = C_1 + C_2\tau^3$. Thus, the tensor-to-scalar evolution variable freezes to a constant value at late times, while the second mode rapidly decays as $\tau\to0^-$. This behavior mirrors the standard freeze-out mechanism of inflationary perturbations and shows explicitly that departures from constant $X_k$ are generated by deviations from exact de Sitter expansion and by differences between the scalar and tensor scattering potentials.

\begin{figure*}[t]
    \centering

    \begin{subfigure}[t]{0.48\textwidth}
        \centering
        \includegraphics[width=\linewidth]{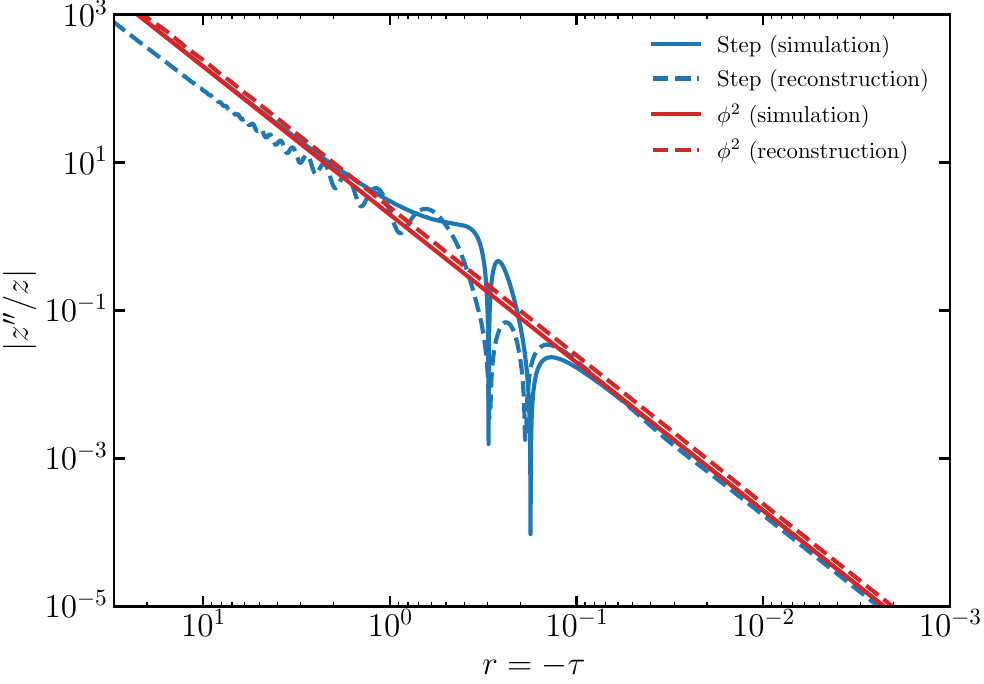}
        \caption{Scalar sector: $|z''/z|$.}
    \end{subfigure}
    \hfill
    \begin{subfigure}[t]{0.48\textwidth}
        \centering
        \includegraphics[width=\linewidth]{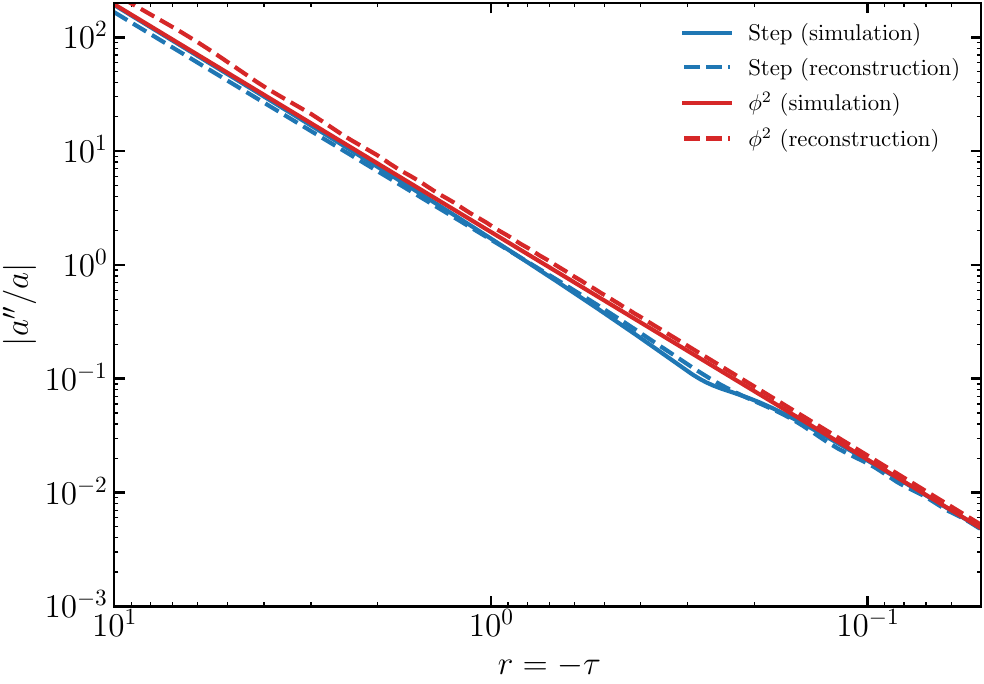}
        \caption{Tensor sector: $|a''/a|$.}
    \end{subfigure}

    \caption{
    \capindent
    Reconstructed and simulated effective potentials for scalar and tensor perturbations in inflation. 
    The left panel shows the scalar quantity $|z''/z|$, while the right panel displays the tensor quantity $|a''/a|$, both as functions of conformal time $r=-\tau$. 
    Solid lines correspond to numerical results obtained from the background evolution, and dashed lines indicate the reconstruction from inverse-scattering methods. 
    Results are shown for a quadratic potential ($\phi^2$) and a step potential. 
    The quadratic model yields a smooth evolution consistent with slow-roll dynamics, whereas the step potential introduces localized features that reflect transient departures from slow-roll behavior. 
    The agreement between simulation and reconstruction demonstrates the ability of the method to capture both the background evolution and localized features in the effective potentials.
    }
    \label{fig:zz_dda_comparison}
\end{figure*}

\section{Large-$k$ Regime and Born Approximation}

The inverse-scattering reconstruction developed in the previous sections can be tested analytically in the short-distance or large-$k$ regime through the Born approximation. In scattering theory, the Born approximation corresponds to the weak-scattering limit, where the phase shift remains small, $|\delta_\ell(k)|\ll1$, such that the scattering matrix becomes
\begin{equation}
    S_\ell(k)
    \simeq
    1-2i\delta_\ell(k).
\end{equation}
In this regime, the Marchenko kernel linearizes and the integral equation simplifies considerably. The transformation kernel satisfies
\begin{equation}
    A_\ell(r,t)\simeq -\mathcal F_\ell(r,t),
\end{equation}
such that the reconstructed potential is approximately given by
\begin{equation}\label{eq:born_reconstruction}
    V_s(r)
    \simeq
    2\frac{d}{dr}\mathcal F_\ell(r,r),
\end{equation}
where $\mathcal F_\ell(r,t)$ is the input kernel defined in Eq.~\eqref{eq:input_kernel}. This approximation provides the simplest analytic realization of the inverse-scattering reconstruction and allows a direct comparison between the reconstructed effective potential and the exact inflationary background evolution.

Physically, the Born regime corresponds to perturbation modes with sufficiently large wavenumbers, where the effective potential acts as a small perturbation over the free oscillatory solution. In the cosmological context, this limit is particularly relevant because future observational probes are expected to extend sensitivity to primordial fluctuations far beyond conventional CMB anisotropy scales. In particular, proposed spectral-distortion missions such as \cite{Kogut:2011xw, PRISM:2013fvg, Chluba:2019nxa, Abitbol:2017vwa} aim to probe primordial fluctuations up to scales of order $k\sim10^3-10^4\ {\rm Mpc}^{-1}$, where departures from canonical slow-roll evolution may leave observable signatures in the primordial spectrum.

\subsection{Benchmark Inflationary Potentials}
\label{sec:working_cases}

To test the inverse-scattering reconstruction framework, we consider two benchmark inflationary models with qualitatively different dynamical behavior.

The first benchmark corresponds to the quadratic potential
\begin{equation}\label{eq:quadratic_potential}
    V(\phi)=m^2\phi^2,
\end{equation}
which provides a smooth slow-roll evolution and generates an approximately scale-invariant primordial spectrum. This model serves as a reference case where the effective potentials $z''/z$ and $a''/a$ evolve smoothly and the inverse-scattering reconstruction is expected to reproduce the background accurately within the Born approximation.

The second benchmark corresponds to a step-like potential,
\begin{equation}\label{eq:step_potential}
    V(\phi) = m^2\phi^n \left[ 1+ \beta \tanh \left( \frac{\phi-\phi_{\rm step}}{\delta} \right) \right],
\end{equation}
where $\beta$, $\delta$, and $\phi_{\rm step}$ characterize the amplitude, width, and location of the step feature, respectively. Such potentials induce transient violations of slow-roll evolution and generate localized oscillatory features in the scalar primordial power spectrum \cite{Starobinsky:1992ts, Adams:2001vc, Palma:2017wxu}. These oscillations arise because the inflaton experiences a sudden change in acceleration while crossing the step, which modifies the effective Mukhanov--Sasaki potential and consequently the scattering properties of the perturbations.

For the numerical implementation shown in this work, we adopt
$n=2$, $\beta=0.20$, $\delta=0.20$, $\phi_{\rm step}=13.5$, with inflaton mass $m_\phi =10^{-7}M_{\rm Pl}$, for both the quadratic and step-potential cases.

For each model, we numerically solve the coupled inflationary background equations, Eqs.~\eqref{eq:KG_f_N} and \eqref{eq:friedmann_H_N}, together with the Mukhanov--Sasaki equation for scalar and tensor perturbations Eq.\eqref{eq:mukhanov_sasaki}. From the background evolution, we compute the exact effective potentials $\frac{z''}{z}$, and  $\frac{a''}{a}$, which provide the expected target reconstruction. In the numerical evaluation of Eq.~\eqref{eq:born_reconstruction}, the Fourier-like integrals defining the input kernel $\mathcal{F}_\ell(r,t)$ are evaluated up to a sufficiently large ultraviolet cutoff $k_{\rm max} \approx \mathcal{O}(10^{4})\,\text{Mpc}^{-1}$, above which the spectra smoothly reach their asymptotic scale-invariant values. This regularizes the high-frequency oscillatory behavior and ensures the numerical stability of the spatial derivative.

Using the resulting primordial scalar and tensor spectra, shown in Fig.~\ref{fig:pps_scalar_tensor}, we then reconstruct the corresponding effective potentials within the Born approximation using Eq.~\eqref{eq:born_reconstruction}. The reconstructed quantities are compared against the exact background evolution in Fig.~\ref{fig:zz_dda_comparison}. This comparison allows us to quantify the validity and limitations of inverse-scattering reconstruction in both smooth, slow-roll backgrounds and models with localized, nontrivial features.

The quadratic potential, therefore, provides a controlled slow-roll benchmark where the Born approximation is expected to perform accurately, while the step potential constitutes a nontrivial test of the reconstruction framework in the presence of transient departures from slow-roll evolution and oscillatory structure in the primordial spectrum.

\subsection{Results}
From Fig.~\ref{fig:zz_dda_comparison}, we can conclude that for the quadratic potential, the reconstructed potentials closely follow the simulated evolution over essentially the entire domain. The reconstructed curves remain nearly parallel to the simulated ones and differ only by a small normalization offset at large $r=-\tau$. This minor discrepancy in the infrared limit ($r \to \infty$) is expected, as it reflects the standard sensitivity of the linearized Marchenko inversion to the low-$k$ truncation of the scattering data, where boundary conditions dominate the integral transformation of the input kernel. This agreement indicates that the Born approximation captures the dominant structure of the effective potential when the inflationary evolution remains close to canonical slow-roll.

The situation changes for the step potential. In the scalar sector, the reconstructed potential reproduces the large-scale behavior of $z''/z$ remarkably well away from the feature region, but localized oscillatory deviations appear around intermediate values of $r$. These deviations coincide with the region where the simulated potential develops sharp transitions induced by the step feature in the inflationary potential. In particular, the exact scalar effective potential exhibits localized oscillations and rapid variations that are only partially recovered by the Born reconstruction.

This behavior is expected because the Born approximation assumes weak scattering and small phase shifts. The step potential generates transient violations of slow-roll evolution, producing localized non-adiabatic effects that enhance mode mixing and induce oscillatory features in the scalar primordial spectrum. Such effects correspond to stronger scattering in the inverse-scattering description and therefore exceed the regime where the linearized Marchenko reconstruction remains fully accurate.

Nevertheless, despite these localized discrepancies, the reconstructed scalar potential still captures the overall envelope and asymptotic scaling of the exact solution. This indicates that the inverse-scattering framework remains sensitive to inflationary features even within the weak-scattering approximation.

The tensor reconstruction is smoother than the scalar reconstruction. The reconstructed tensor potential $a''/a$ closely follows the simulated evolution for both the quadratic and step models over nearly the entire range shown. This difference reflects the fact that tensor perturbations probe only the background geometry through $a''/a$, whereas scalar perturbations depend on the combination $z''/z$, which is directly sensitive to inflaton dynamics and transient slow-roll violations.

Consequently, the tensor sector appears substantially less sensitive to localized features in the inflationary potential. The Born approximation therefore performs better for tensor reconstruction because the corresponding effective scattering potential remains smoother and closer to the weak-scattering regime.

Overall, Fig.~\ref{fig:zz_dda_comparison} demonstrates that the inverse-scattering reconstruction successfully recovers the dominant behavior of the inflationary effective potentials from the primordial spectra. The method performs particularly well for smooth slow-roll backgrounds, while localized deviations from slow-roll introduce oscillatory corrections that partially challenge the validity of the Born approximation. The comparison also highlights the complementary role of scalar and tensor perturbations: scalar reconstruction is more sensitive to detailed inflaton dynamics, whereas tensor reconstruction primarily traces the background expansion history.

\section{Conclusions}\label{sec:conclusion}
In this work, we explored the use of inverse-scattering techniques as a reconstruction framework for inflationary perturbations. By rewriting the Mukhanov--Sasaki equation in a Schrödinger-like form, we related the effective inflationary potentials, $z''/z$ for scalar modes and $a''/a$ for tensor modes, to scattering data encoded in the Jost function. This provides a direct connection between the primordial power spectra and the reconstructed background quantities governing the evolution of cosmological perturbations.

A central result of the analysis is that the freeze-out amplitude of the Mukhanov--Sasaki variable is naturally associated with the Jost solution rather than with the regular solution. The Jost solution is fixed by the Bunch--Davies condition in the sub-horizon regime, and its small-$r$ behavior determines the observable super-horizon amplitude. In contrast, the regular solution is proportional to the decaying mode and therefore does not directly represent the physical growing perturbation relevant for the primordial spectrum.

We derived the scalar and tensor primordial spectra in terms of their corresponding Jost functions, $F^{(s)}_{\nu_s-1/2}(k)$ and $F^{(t)}_{\nu_t-1/2}(k)$. This allowed us to express the tensor-to-scalar ratio as a ratio of scalar and tensor Jost amplitudes. In the slow-roll limit, where the Jost functions reduce to pure phases and $\nu_s\simeq\nu_t\simeq3/2$, the standard consistency relation $\mathbf r\simeq16\epsilon=-8n_t$ is recovered. We also introduced the diagnostic variable $X_k=v_k/u_k$, which describes the relative evolution between tensor and scalar perturbations and is sourced by the difference $(z''/z-a''/a)$.

To test the reconstruction, we implemented the Born approximation in the large-$k$ regime. This approximation linearizes the Marchenko equation and provides a controlled weak-scattering limit. We applied the method to two benchmark inflationary models: a smooth quadratic potential and a step potential. The quadratic potential serves as a slow-roll reference case, whereas the step potential introduces transient departures from slow-roll and generates localized features in the primordial spectra.

The comparison between simulated and reconstructed effective potentials shows that the Born approximation accurately captures the dominant behavior of both $z''/z$ and $a''/a$ for the quadratic model. For the step potential, the reconstruction still reproduces the global scaling and asymptotic behavior, but localized discrepancies appear in the scalar sector near the feature region. These discrepancies reflect the breakdown of the weak-scattering assumption in the presence of transient non-adiabatic dynamics. The tensor reconstruction remains smoother and more accurate, consistent with the fact that tensor modes probe only the background geometry through $a''/a$, while scalar modes are directly sensitive to inflaton dynamics through $z''/z$.

Our results show that inverse scattering provides a useful and physically transparent framework for reconstructing inflationary dynamics from primordial spectra. The method is particularly robust for smooth slow-roll backgrounds and remains sensitive to localized features even when the Born approximation is only partially valid. Future work should go beyond the Born approximation by solving the full Marchenko equation, incorporating finite observational windows in $k$-space, and exploring non-canonical scenarios where the scalar and tensor sectors may encode additional information about the sound speed or higher-derivative interactions. Looking forward, the inverse-scattering formalism developed in this work opens an exciting avenue for probing the inflationary potential at small cosmological scales. The high-$k$ regime ($k \sim 10^3 - 10^4\,\text{Mpc}^{-1}$), where the linearized Born approximation exhibits its maximum numerical stability and accuracy, is precisely the window that will be targeted by future high-precision cosmological observations. Specifically, upcoming CMB spectral distortion missions, such as PIXIE and proposed successor experiments, along with next-generation Large-Scale Structure (LSS) surveys, will provide unprecedented constraints on the primordial power spectra at these scales. By mapping potential features directly from future data back to the inflationary action, this inverse-scattering framework stands as a powerful, model-independent tool to reconstruct the fine structure of the early universe.

\begin{acknowledgments} 
\noindent JM wish to acknowledge the support of the program \textit{Investigadores e Investigadoras por M\'exico} of SECIHTI.
\end{acknowledgments}

\appendix

\section{Power Spectrum of the Regular Solution}\label{appendix:regular_solution} The scalar primordial power spectrum, written in terms of the regular solution becomes 
\begin{equation}
\begin{split}
\mathcal P_s^{(\mathrm{reg})}(k)
&=
\frac{1}
{2^{2\nu_s+4}\pi\Gamma^2(\nu_s+1)
\left|F^{(S)}_{\nu_s-\frac12}(-k)\right|^2}
\\
&\quad\times
\frac{(1-\epsilon_1)^{-2\nu_s-1}}{\epsilon_1}
\frac{H^2}{M_{\rm Pl}^2}
\left(\frac{k}{aH}\right)^{2\nu_s+3}.
\end{split}
\end{equation}
A crucial conceptual point is that the physical freeze-out mode behaves as $ u_k(\tau)\propto(-\tau)^{\frac12-\nu_s}$, which corresponds to the growing mode associated with the Jost solution. By contrast, the regular solution is proportional to the decaying mode,  $u_k(\tau)\propto(-\tau)^{\nu_s+\frac12}$, and therefore does not directly correspond to the observable cosmological perturbation.

\section{Additional Relations for the Tensor-to-Scalar Ratio}\label{appendix:additional_relations}

In this appendix, we briefly discuss two auxiliary relations that can provide additional insight into the inverse-scattering formulation.

Using the first-order slow-roll relations Eq.~\eqref{eq:scalar_effective_potential}, one may formally solve for the slow-roll parameter $\epsilon$, and substituting into the leading-order consistency relation Eq.~\eqref{eq:r_slow_roll} yields
\begin{equation}
    \mathbf r(k_*) \simeq \frac{16}{3} \left[ \tau_*^2 \left.\frac{z''}{z}\right|_{\tau_*} - \left( 2+\frac32\eta_* \right) \right] \,,
\end{equation}
where all quantities with $_*$ are evaluated at horizon-crossing time $\tau_*$ corresponding to the pivot scale $k_*$ through $k_* = 0.05\,\mathrm{Mpc}^{-1}$.

Introducing the logarithmic variable $Y_k(\tau) \equiv \ln X_k(\tau)$, from Eq.~\eqref{eq:Xk_equation} one obtains
\begin{equation}
    Y_k''+(Y_k')^2+2\frac{u_k'}{u_k}Y_k'
+\left(\frac{z''}{z}-\frac{a''}{a}\right)=0.
\end{equation}
This Ricatti-type equation provides an alternative description of the relative evolution between the ratio of tensor and scalar perturbations. Since the quantity $Y_k$ measures the logarithmic growth of the tensor-to-scalar ratio, the equation makes explicit how derivations between the scalar and tensor effective potentials source departure from the constant canonical value. This formulation can also be used in future WKB analyses and phase-shift formulations of inverse-scattering problems.

\bibliography{apssamp}

\end{document}